\documentclass[aps, showkeys,nofootinbib,floatfix]{revtex4}

\usepackage{amssymb}
\usepackage{amsmath}
\usepackage{graphicx}
\usepackage{hyperref}
\usepackage{graphicx}

\begin{document}

\title{The Fourth Gravity Test and Quintessence Matter Field}
\author{Molin Liu$^{1}$}
\email{mlliu@mail2.xytc.edu.cn}
\author{Benhai Yu$^{1}$}
\author{Fei Yu$^{2}$}
\author{Yuanxing Gui$^{2}$}
\email{guiyx@dlut.edu.cn}
\affiliation{$^{1}$College of Physics and Electronic Engineering,
Xinyang Normal University, Xinyang, 464000, P. R. China\\
$^{2}$School of Physics and Optoelectronic Technology,
Dalian University of Technology, Dalian, 116024, P. R. China}

\begin{abstract}
After the previous work on gravitational frequency shift, light deflection (arXiv:1003.5296) and perihelion advance (arXiv:0812.2332), we calculate carefully the fourth gravity test, i.e. radar echo delay in a central gravity field surrounded by static free quintessence matter, in this paper. Through the Lagrangian method, we find the influence of the quintessence matter on the time delay of null particle is presence by means of an additional integral term. When the quintessence field vanishes, it reduces to the usual Schwarzschild case naturally. Meanwhile, we also use the data of the Viking lander from the Mars and Cassini spacecraft to Saturn to constrain the quintessence field. For the Viking case, the field parameter $\alpha$ is under the order of $10^{-9}$.  However, $\alpha$ is under $10^{-18}$ for the Cassini case.
\end{abstract}



\maketitle

\section{Introduction}
The discovery of dark energy is a landmark of cosmology. In 1998 \cite{Riess}, it became known that our universe is accelerating according to the observation data from the distant supernova. This fact was proved later by further Ia Supernovae (SNe Ia) \cite{Ia} and microwave background (CMB) \cite{CMB} of WMAP. The special matter contained a negative pressure which can drive the universe acceleration, called dark energy. The large scale distribution of galaxies observations \cite{largescalar} also tells us there is about 70\% dark energy in the universe, which seems more like an evenly distributed background component and it cannot assemble in the large scale. Else, the theories of big bang nucleosynthesis and galaxy formation confine in the early universe the ratio of dark energy to very small values and only it becomes large after galaxy formation. In the theoretical analysis, the simplest candidate is the cosmological constant. If so, our universe will be further accelerating and the whole space will more like a de Sitter topological structure. The cosmological constant combined to cold dark matter leads to a so-called $\Lambda$CDM cosmological model \cite{cosconstant}. In spite of the success of $\Lambda$CDM model in many respects, there are still two big troubles in this model; i.e. the famous fine-tuning and cosmic coincidence problems. Hence, many theoretical physicists prefer a dynamical dark energy model to a $\Lambda$CDM model. The simplest dynamical model is a quintessence field which is a scalar field with canonical momentum \cite{q}. When quintessence rolls along itself the potential curve, it will provide a negative pressure if the potential curve is very flat i.e. the quintessence field is at a slow roll state. Else, its attractor solution can solve the cosmic coincidence very well.

As we know that, the cosmological constant model can give a static spherically symmetric black hole i.e. Schwarzschild-de Sitter black hole (SdS-BH) which is corresponding to a usual Schwarzschild case. However, how to obtain a counterpoint to that of Schwarzschild case just like the SdS-BH in cosmological constant model? After this problem occurred some fundamental work had been done by people such as Gonzalez\cite{Gonzalez}, Chernin et.al.\cite{Chernin} and Kiselev \cite{Kiselev} and so on. In the early works, there were no horizon and no `hair' problems \cite{Gonzalez, Chernin} which means no black hole could be embraced into these metrics. Later, in order to solve this problem Kiselev employed nonzero off-diagonal energy momentum tensor in which its coefficients satisfied additivity and linearity conditions. Hence, a steady black hole solution is obtained \cite{Kiselev}. The process is introduced simply in part II.

Since Kiselev \cite{Kiselev} obtained the exact black hole solutions encoded in quintessence, many people have performed studies on its special properties including: quasinormal frequencies (QNM) \cite{quasinormal},
thermodynamics entropy \cite{entropy}, gravity geodesic precession\cite{geodetic} and the classical three gravity tests on solar system \cite{Liu1122,Liu1123} and so on. The last phenomenology aspects refer to the gravitational frequency shift, the deflection of light and the precession of perihelion of Mercury.
In Ref.\cite{Liu1122} via calculating the photon's 4D momentum, it is found that the photon frequency increases for the range $\omega_q \in [-1, -1/3]$ and decreases for the range $[-1/3, 0]$, where $\omega_q$ is the quintessence state parameter satisfying the equation of sate (EOS). The field parameter $\alpha$ is sensitively dependent on the state parameter $\omega_q$.  Comparing with the H-masers of the GP-A redshift experiment data, the constraint on quintessence matter is $\alpha \in (10^{-28}, 10^{-7})$ with the range $\omega_q \in [-1, 0]$. Meanwhile, for the light deflect test its influence behaviors also are different according to various $\omega_q$ values. The analytic results are obtained via the integral photon's Binet equation with special solvable value $\omega_q = \{-1, -2/3, -1/3, 0 \}$. By using the long-baseline radio interferometry data, the corresponding constraints on the quintessence field parameter $\alpha = \{-, 10^{-34}, 10^{-3}, 10^{27} \}$ are obtained where the label ``$-$" means that there is no influence on the deflection of light by quintessence for the case $\omega_q = -1$ \cite{Liu1122}. Then after that, the test of perihelion advance has been presented in Ref. \cite{Liu1123}. The trajectory of the test mass and the motion of a binary system are discussed in detail. However, as far as we know there is no work relating to ``the fourth test" \cite{shapiro,Willtest} in gravity field i.e. time delay in gravity test. Considering the above various factors, we calculate carefully the behavior of gravity time delay influenced by quintessence matter and try to find the constraints on quintessence field parameter in alternative view here.

This paper is organized as follows: in section II, we present the
Kiselev solution briefly. In section III, we use the Lagrangian method to find the analytic expression of time delay in different points. In section IV,
we use the data of the Viking lander on Mars and the Cassini spacecraft on the way to Saturn to constrain the quintessence field parameter. Section V is the conclusion. We adopt the signature $(+, -, -, -)$ and put $\hbar$, $c$, and $G$ equal to
unity.

\section{The static spherically symmetric black hole solutions with free quintessence matter}
The modified Schwarzschild space surrounded by static spherically symmetric quintessence matter is given by \cite{Kiselev}
\begin{equation}\label{metricq}
   d s^2 = \bigg{(}1 - \frac{2M}{r} - \frac{\alpha}{r^{3\omega_q +
    1}}\bigg{)} d t^2 - \bigg{(}1 - \frac{2M}{r} - \frac{\alpha}{r^{3\omega_q +
    1}}\bigg{)}^{-1} d r^2 -r^2 (d \theta^2 + \sin^2\theta d
    \varphi^2),
\end{equation}
where $M$ is the black hole mass, $\omega_q$ is the state parameter of quintessence, $\alpha$ is the corresponding quintessence field parameter. In this paper, we evaluate the quintessence field with $\omega_q$ in the range of $[-1, 0]$ without considering the supernova dimming. Interestingly, this metric reduces to the pure Schwarzschild space and the Schwarzschild-de Sitter space for the cases of $\alpha = 0$ and $\omega_q = -1$, respectively. The energy momentum coupling with evenly distributed quintessence is shown as
\begin{eqnarray}
  T_{t}^{t} &=& \rho_q (r), \label{tt}\\
  T_{i}^{j} &=& \rho_q (r) \gamma \left[-(1 + 3B)\frac{r_{i}r^{j}}{r_{n}r^{n}} + B
  \delta_{i}^{j}\right],\label{ij}
\end{eqnarray}
where $\rho_q$ is the density of quintessence matter.
The internal structure $B$ function satisfied $ B = - \left(3\omega_q + 1\right)/6\omega_q$ via the additivity and linearity principle $T_{t}^{t} = T_{r}^{r}$. The isotropic
average over the angle components is
\begin{equation}\label{average}
    <T_{i}^{j}> = -\rho_{q}(r)\frac{\gamma}{3}\delta_{i}^{j} =
    -p_{q}(r) \delta_{i}^{j},
\end{equation}
where the relationship $<r_{i}r^{j}> =
\frac{1}{3}\delta_{i}^{j}r_{n}r^{n}$ is used and the state equation
is $P_{q} = \omega_q \rho_q$ where $\omega_q = \gamma/3$.

With the relationship $\lambda
+ \nu = 0$, the energy momentum tensor components are
\begin{equation}
  T_t^t = T_r^r = \rho_q; \ \ T_{\theta}^{\theta} = T_{\varphi}^{\varphi} =
  -\frac{1}{2}\rho_q (3\omega_q + 1). \label{theta}
\end{equation}
The quintessence density is
\begin{equation}\label{rho}
 \rho_q =
\frac{\alpha}{2}\frac{3\omega_q}{r^{3(1+\omega_q)}}.
\end{equation}
So if we require the density of energy positive, $\rho_q > 0$, we
deduce that $\alpha$ is negative for $\omega_q$ negative. The
curvature has the form of
\begin{equation}\label{curvature}
    R = 2 T_{\mu}^{\mu} = 3 \alpha \omega_q \frac{1-3\omega_q}{r^{3(\omega_q +
    1)}}.
\end{equation}
Apparently, $r = 0$ is the singularity for $\omega_q\ \neq \{-1,\ 0,\
1/3 \}$. The free quintessence
creates an outer horizon of de Sitter universe at $r = r_q$ for
\begin{equation}\label{outerhorizon}
    -1 < \omega_q < -\frac{1}{3},
\end{equation}
and also generates an inner horizon of black hole at $r = r_q$ for
\begin{equation}\label{innerhorizon}
    -\frac{1}{3} < \omega_q <0.
\end{equation}

\section{radar echo delay on Solar system}
Except for the gravitational redshift \cite{Liu1122}, perihelion procession \cite{Liu1123} and bending of light \cite{Liu1122}, a further measurable effect concerning the null geodesics is the ``time delay" of radar signals. Unlike the flat space, the travel time of light between any two given points increases due to the presence of space curvature. This type of travel increasing time can be measured via using the propagation of radar signals in solar system, which is proposed firstly by Shapiro in 1964 \cite{shapiro}. Else, it is also constituted by the ``fourth test" of GR known as radar ranging \cite{Willtest}. Here, the time delay for radar ranging is calculated in the space of metric (\ref{metricq}) by the Lagrangian method. The Lagrangian for a test mass particle in the field described by the equation (\ref{metricq}) is
\begin{equation}\label{lagrangian}
    \mathcal{L}_p = \left[- f(r) \dot{t} + \frac{\dot{r}}{f(r)} + r^2 \left(\dot{\theta} +\sin^2 \theta \dot{\varphi}\right)\right]^{1/2},
\end{equation}
where the overdot represents differentiation with respect to the affine parameter $\xi$ along the geodesics. The Euler-Lagrange equations read as follows
\begin{equation}\label{ELequations}
    \frac{d}{d \xi} \left(\frac{\partial \mathcal{L}_p}{\partial \dot{x}^{\mu}}\right) - \frac{\partial \mathcal{L}_p}{\partial x^{\mu}} = 0,
\end{equation}
Without loss of generality, the observer is confined to orbits with $\theta = \pi/2$ and $\dot{\theta} = 0$. Hence, the Lagrangian $\mathcal{L}_p$ becomes
\begin{equation}\label{lagrangian0}
    \mathcal{L}_p = \left(- f(r) \dot{t} + \frac{\dot{r}}{f(r)} + r^2 \dot{\varphi}\right)^{1/2}.
\end{equation}
According to the cyclic coordinates $\varphi$ and $t$, we can identify two constants of motion,
\begin{eqnarray}
   && r^2 \frac{d\varphi}{d \zeta} \label{m1} = L,\\
   && (1 - \frac{2M}{r} - \frac{\alpha}{r^{3\omega_q + 1}}) \frac{d t}{d
   \zeta} = E\label{m2}.
\end{eqnarray}
The third motion equation can be given by the normalization relation
of photons $g_{\mu\nu} p^{\mu}p^{\nu}$ = 0,
\begin{equation}\label{m3}
    \left(\frac{d r}{d \zeta}\right)^2 = E^2 - \frac{L^2}{r^2}\left(1 - \frac{2M}{r} - \frac{\alpha}{r^{3\omega_q +
    1}}\right).
\end{equation}
The dynamic evolution of photon is determined fully by
Eqs.(\ref{m1}), (\ref{m2}) and (\ref{m3}) in this space. Combining
Eq. (\ref{m1}) with (\ref{m3}), the photon's trajectory equation can
be gotten as
\begin{equation}\label{trajectoryequation}
    \left(\frac{1}{r^2}\frac{d r}{d \varphi}\right)^2 =
    \left(\frac{E}{L}\right)^2 - \frac{1}{r^2} \left( 1- \frac{2M}{r} - \frac{\alpha}{r^{3 \omega_q +
    1}}\right).
\end{equation}
Meanwhile, we define two new parameters, one is the impact parameter
$b = L/E$, which means the effective sighting range, the other is
the photon's effective potential $1/B^2(r)$ where
\begin{equation}\label{ep}
    B (r) = r \left(1 - \frac{2M}{r} - \frac{\alpha}{r^{3 \omega_q + 1}}
    \right)^{-1/2}.
\end{equation}

\begin{figure}
  \includegraphics[width=3.5 in]{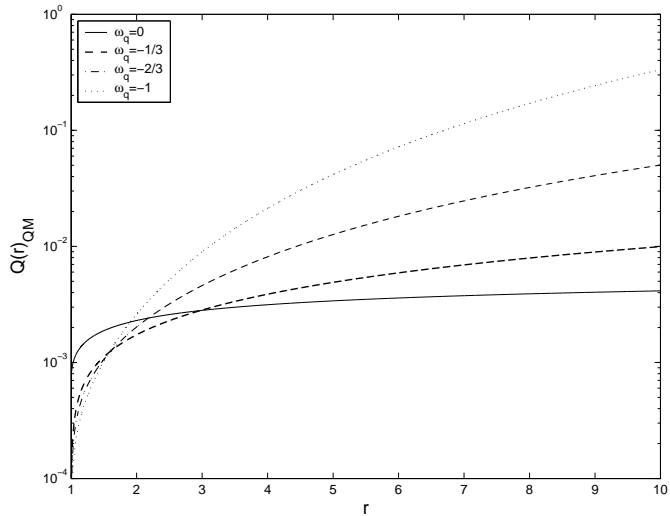}\\
  \caption{Additional term $t(r,r_0)_{QM}$ versus radial $r$ coordinates for $\alpha = 10^{-3}$ and solar radius $r_0 = 1$.}\label{fig1}
\end{figure}

According to Eq. (\ref{ep}), the photons' orbital equation
(\ref{trajectoryequation}) can be rewritten as
\begin{equation}\label{or3}
    \left(\frac{1}{r^2} \frac{d r}{d \varphi}\right)^2 =
    \frac{1}{b^2} - \frac{1}{B^2(r)}.
\end{equation}
The distance of closest approach $r_0$ is defined by
\begin{equation}\label{closestappro}
    \frac{d r}{d \varphi}|_{r=r_0} = \pm r_0^2 \left(\frac{1}{b^2} - \frac{1}{B^2(r_0)}\right)^{1/2}
\end{equation}
So we can get the relationship $b = B(r_0)$. According to Eqs.(\ref{m2}) and (\ref{m3}), the differential relation between time $t$ and radius $r$ is obtained as follows:
\begin{equation}\label{timeandradius}
    \frac{d t}{d r} = \pm \frac{1}{b f(r)}\left(\frac{1}{b^2} - \frac{1}{B^2(r)}\right)^{-1/2},
\end{equation}
where $f (r) = 1 - 2M/r - \alpha/r^{3\omega_q + 1}$.
So the time travel for the null geodesic is
\begin{equation}\label{timetravel}
   \nonumber t(r,r_0) = \int_{r_0}^{r}\frac{d r}{f(r)} \left[1 - \frac{f(r)}{f(r_0)} \frac{r_0^2}{r^2}\right]^{-1/2} = t(r,r_0)_{GR} + t(r,r_0)_{QM},
\end{equation}
where
\begin{eqnarray}
t(r,r_0)_{GR} &=& \int_{r_0}^{r} d r \left(1 - \frac{r_0^2}{r^2}\right)^{-1/2} \left(1 + \frac{M r_0}{r (r + r_0)} + \frac{2M}{r}\right), \label{GR1}\\
t(r,r_0)_{QM} &=& \int_{r_0}^{r} d r \left(1 - \frac{r_0^2}{r^2}\right)^{-1/2}\left[\frac{r^{3\omega_q + 1} - r_0^{3\omega_q + 1}}{r_0^{3\omega_q - 1} r^{3\omega_q + 1} \left(r^2 - r_0^2\right)} + \frac{2}{r^{3\omega_q + 1}}\right] \frac{\alpha}{2}. \label{QM2}
\end{eqnarray}
The first term $t(r,r_0)_{GR}$ is the usual case in GR \cite{book} and the last one is the effect of quintessence field which will be solved particular in the remained parts. Apparently, once the quintessence field disappears, the result is reduced to the Schwarzschild one. Because the undetermined state parameter $\omega_q$ which is coupling in the exponent of $r$ or $r_0$, we select the solvable integers $3 \omega_q + 1 = (-2,\ -1,\ 0,\ 1)$ with $\omega_q \in [-1, 0]$ to integrating Eq. (\ref{QM2}). This kind of method, which also is shown in Ref. \cite{Liu1122} and \cite{Liu1123}, is useful to solve approximately sophisticated differential equation in which the independent variable contains some unknown parameter. The integrating results are shown in TABLE \ref{table1}. The corresponding sketchy outline of the additional term $t(r,r_0)_{QM}$ versus radial $r$ is shown in Fig. (\ref{fig1}). It illustrates clearly that the influence of quintessence on time delay heightens with the decreasing quintessence state parameter $\omega_q$ correspondingly.
\begin{table*}[!h]\scriptsize
\caption{The forms of $t(r,r_0)_{QM}$} \label{table1}
\begin{center}\normalsize
\begin{tabular}{l|l}
     \hline
            \ \ $\omega_q$ \ \ \parbox[][3em][c]{20em}{}& $t(r,r_0)_{QM}$ \\
     \hline
      \ \ 0 \ \ \  \parbox[][3em][c]{20em}{}& $\frac{\alpha}{2} \left(1 - \frac{r_0^2}{r^2}\right)^{-1/2} \left[1 - \frac{r_0}{r} + 4\sqrt{1-\frac{r_0}{r}}\sqrt{1+\frac{r_0}{r}}\log \left(\sqrt{r-r_0} + \sqrt{r + r_0}\right)\right]$\\
      \ \ -1/3 \ \ \ \ \ \parbox[][3em][c]{20em}{}& $\alpha r \sqrt{1 - \frac{r_0^2}{r^2}}$\\
      \ \ -2/3 \ \ \  \parbox[][3em][c]{20em}{}& $\frac{\alpha}{2 r} \left(r^3 - r_0^3\right)\left(1 - \frac{r_0^2}{r^2}\right)^{-1/2}$\\
      \ \ -1 \ \ \  \parbox[][3em][c]{20em}{}& $\frac{\alpha r}{6} \left(2 r^2 + r_0^2\right)\sqrt{1 - \frac{r_0^2}{r^2}}$\\
\hline
\end{tabular}
\end{center}
\end{table*}

\section{constraints from experiment of Viking lander on Mars}
The time interval between emission and return receiver is
\begin{eqnarray}
    \Delta T &=& 2 t(r_{\oplus}, r_0) + 2 t(r_{R}, r_0)\\
    &=& \Delta T_{GR} + \Delta T_{\alpha},\label{tinterval}
\end{eqnarray}
where the first term $\Delta T_{GR}$ is the usual GR value the modified term $\Delta T_{\alpha}$ depends on the state parameter $\omega_q$ of quintessence which is presented in TABLE \ref{table2} with the limits of $r_{\oplus} \gg r_0$ and $r_{R}\gg r_0$. The proper time $\Delta \tau$ which elapses on Earth is in relation to the change in coordinate time $\Delta T$ in the form $\Delta \tau = \sqrt{g_{00}(r_\oplus)} \Delta T$.
\begin{table*}[!h]\scriptsize
\caption{The estimates on quintessence field parameter $\alpha$ from Mars} \label{table2}
\begin{center}\normalsize
\begin{tabular}{l|c|l}
     \hline
            $\omega_q$ & $\Delta T_{\alpha}$& Estimates on \\
     \hline
     0 \ \ \  \parbox[][3em][c]{20em}{}& $\alpha \left(2 + 4\log 4 \sqrt{r_{R} r_{\oplus}}\right)$ \ \ \ \ & $|\alpha| \leq 4.52902 \times 10^{-9}$ \\
     -1/3 \ \ \  \parbox[][3em][c]{20em}{}& $2 \alpha \left(r_{\oplus} + r_{R}\right)$ & $|\alpha| \leq 5.9761 \times 10^{-19}$\\
     -2/3 \ \ \  \parbox[][3em][c]{20em}{}& $\alpha \left(r_{\oplus}^2 + r_{R}^2\right)$&$|\alpha| \leq 3.10542 \times 10^{-41}$\\
     -1 \ \ \  \parbox[][3em][c]{20em}{}& $\frac{2 \alpha}{3} \left(r_{\oplus}^3 + r_{R}^3\right)$& $|\alpha| \leq 3.78879 \times 10^{-41}$\\
\hline
\end{tabular}
\end{center}
\end{table*}

So the excess round trip $\Delta \tau$ for signal emitted from Earth and bounced off the other planet can be obtained as
\begin{equation}\label{roundtrip}
    \Delta \tau = \Delta \tau_{GR} \left(1 + \Delta_{TD}\right)
\end{equation}
The GR value $\Delta \tau_{GR}$ is
\begin{equation}
\Delta \tau_{GR} = 4 M \left[\log \left(\frac{r_\oplus + \sqrt{r_\oplus^2 - r_0^2}}{r_0}\right) + \log \left(\frac{r_R + \sqrt{r_R^2 - r_0^2}}{r_0}\right)\right]
\approx 4 M \log \frac{4 r_\oplus r_R}{r_0^2} \approx 2.40\times10^{-4} s. \label{deltagr}
\end{equation}
The best experimental constraints from Solar system on the time delay is the Viking lander on Mars and give $|\Delta_{TD}| \leq 0.002$ \cite{Reasenberg}. The distances of Earth and Mars from Sun are $r_\oplus = 1.525 \times 10^{13} cm$ and $r_R = 2.491 \times 10^{13} cm$, and the Sun radius is $r_0 = 6.955 \times 10^{10} cm$. The final additional time delay $\Delta T_{\alpha}$ caused by quintessence field is shown in Table \ref{table2}. Meanwhile, we also give the corresponding constraints on field parameter $\alpha$ in TABLE \ref{table2} based on the data of Viking lander on Mars \cite{Reasenberg}. Else, we can read that the smaller the parameter $\omega_q$, the stronger constraints on $\alpha$.

\section{constraints from Cassini spacecraft experiment}
It is also known that the highest precision data actually have come from the tracking of spacecraft which emit signal rather than reflection of radar off of planets for its surface influence \cite{book,Reasenberg}. Here we use the data of Cassini spacecraft from Cassini-Huygens mission which are composed of two main elements: the NASA Cassini orbiter and the ESA Huygens probe. The changed frequencies for Cassini spacecraft experiment are \cite{SdS}
\begin{equation}\label{changedfrequency}
    y = \frac{\nu (t) - \nu (0)}{\nu (0)} = \frac{d}{d t} \Delta T
\end{equation}
where $\nu (0)$ is the emitted frequency from Earth and $\nu (t)$ is the received frequency by Earth, respectively. The contribution to frequencies by quintessence, i.e. $\alpha$-term, is
\begin{equation}\label{alphaterm}
    y_{\alpha} = \frac{d}{d t} \Delta T_{\alpha}.
\end{equation}
It is known that the time delay experiment is measured nearly 25 days i.e. 12 days before conjunction and 12 days after. The Cassini, the Sun and the Earth are almost aligned at conjunction and the geocentric distance of spacecraft is 8.43 AU, with a minimum impact parameter $b_{min} = 1.6 R_{\odot}$. According to these measures, the general relativity ($\lambda = 1$) contribution to $y(\nu)$ is $6\times 10^{-10}$ within an accuracy of $10^{-14}$. This leads immediately to a upper bound
\begin{equation}\label{frequencyalpha}
    y_{\alpha} (12d) - y_{\alpha} (0) \leq 10^{-14}.
\end{equation}
Because of the spacecraft being much farther away from Sun than Earth, we adopt that $r_0'=d r_0/d t \approx v_{\oplus} \approx 29.78\ Km/sec$ is approximately Earth's orbiting velocity around Sun. For the distance of Cassini is adopted approximately its conjunction position i.e. $r_R = 8.43 AU - 1 AU = 1.1 \times 10^{12} m$. According to the various quintessence state parameters $\omega_q$ and the above various conditions, four types of additional frequency $y_{\alpha}$ are listed in Table III. Except for the case of $\omega_q = 0$, the trends of constraints on $\alpha$ versus $\omega_q$  are similar with the experiment of the Viking lander on Mars.
\begin{table*}[!h]\scriptsize
\caption{The estimates on quintessence field parameter $\alpha$ from Cassini spacecraft experiment} \label{table3}
\begin{center}\normalsize
\begin{tabular}{l|c|l}
     \hline
            \ \  $\omega_q$ &\ \ \  $y_{\alpha}$&\ \ Estimates on\\
     \hline
      \ \ 0 \parbox[][3em][c]{20em}{} \ \ \  & $\alpha r_0 r_0' \left[\left(\frac{1}{r_{\oplus}^2} + \frac{1}{r_R^2}\right)^2 - \frac{1}{r_0} \left(\frac{1}{r_{\oplus}} +\frac{1}{r_R}\right)+\frac{2}{r_0 r_0'}\left(r_{\oplus} + r_R\right)\right]$ & $|\alpha|\leq 3.3344\times 10^{-28}$\\
      \ \ -1/3\parbox[][3em][c]{10em}{} \ \ \ \ \ & $-2 \alpha r_0 r_0' \left(\frac{1}{r_{\oplus}} + \frac{1}{r_R}\right)$ & $|\alpha|\leq 2.6473\times 10^{-18}$\\
      \ \ -2/3\parbox[][3em][c]{10em}{} \ \ \  & $\alpha r_0 r_0' \left[-3 r_0 \left(\frac{1}{r_{\oplus}} +\frac{1}{r_R}\right) + 2\right]$& $|\alpha|\leq 2.0263\times 10^{-29}$\\
      \ \ -1\parbox[][3em][c]{10em}{} \ \ \  & $-\frac{\alpha r_0^3 r_0'}{3} \left(\frac{1}{r_{\oplus}} +\frac{1}{r_R}\right)$ & $|\alpha|\leq 3.279\times 10^{-35}$\\
\hline
\end{tabular}
\end{center}
\end{table*}
\section{conclusion}
In this paper, we have used the Lagrangian formulation to analyze the influence of quintessence field on ``the fourth gravity test" i.e. radar echo delay in a central gravity field without considering the internal charge. By using the data of Viking lander and Cassini mission, we have obtained the constraints of quintessence field from alternative channels. Now, two points need to be emphasized especially as follows.

1. There are two key parameters $\omega_q$ and $\alpha$ in this model. The formal $\omega_q$, which connects the pressure $p_q$ and energy density $\rho_q$ via the state equation $p_q = \omega_q \rho_q$, is the state parameter of the quintessence. The latter $\alpha$ is the quintessence field parameter, which indicates the influence of quintessence matter over the space outside the black hole. Here we need to pay more attention on the cause and effect of $\alpha$. In Kiselev's original work \cite{Kiselev}, one additively and linearity condition $T^t_t = T^r_r$ is adopted to solve the previous ``no hair no horizon" problem \cite{Gonzalez,Chernin}. Combining the Einstein equation in spherically symmetric static space, the quintessence energy momentum tensors (\ref{tt}) and (\ref{ij}) are finally reduced into a differential equation,
\begin{equation}\label{fequation}
   r^2\frac{d^2 f}{d r^2} + 3 (1 + \omega_q) r \frac{d f}{d r} + (3\omega_q + 1) f= 0,
\end{equation}
where function $f$ is decided by metric function $e^\lambda$ via $\lambda = -\ln (1 + f)$. Then Eq.(\ref{fequation}) has two solutions,
\begin{eqnarray}
  f_q &=& \frac{\alpha}{r^{3\omega_q + 1}}, \label{fq} \\
  f_{BH} &=& -\frac{r_g}{r}. \label{fbh}
\end{eqnarray}
Here, $\alpha$ and $r_g$ are the normalization factors. $r_g$ is the usual Schwarzschild solution and $\alpha$ represents the quintessence field parameter. The expression (\ref{fq}) is the original position where parameter $\alpha$ appear firstly in this model. It is helpful to understand its meaning from two aspects. One is the density of the quintessence energy $\rho_q = 3 \alpha \omega_q/2 r^{3(1+\omega_q)}$. So the sign of normalization constant $\alpha$ and the matter
state parameter $\omega_q$ have to satisfy the condition of $\alpha \omega_q \geqslant 0$. Considering the fixed state parameter, the quintessence increases with bigger $\alpha$. Hence, its magnitude manifests enough the strength of quintessence field. The other feature is the curvature of space $R = 3 \alpha \omega_q (1-3\omega_q)/r^{3(\omega_q + 1)}$. If one fixes the state parameter, we can find the curvature $R$ will be larger with bigger $|\alpha|$ where this point  also means the space will be bended more. From the expressions of density we can find that the dimension of $\alpha$ is sensitively
dependent on the value of state parameter $\omega_q$. The specific dimensional form of $\alpha$ is expressed in Table 6.1 in Ref. \cite{Liu1122} according to various state parameters. If the quintessence field could be treated as the dark energy and the critical density of the universe could be assumed as the value of $\rho_q$, we could find that with bigger astronomical scales the magnitude of $|\alpha|$ is larger. In another words, the quintessence field can have influence more deeply on the large astronomical scale.

2. In this model, we only consider a neutral black hole as well as SdS-BH. The internal charge is not involved in this paper. If we want to consider the charged case, it should refer to the general solution for the Reissner-Nordstr$\ddot{o}$m-de Sitter black hole surrounded by the quintessence matter \cite{Kiselev}, which is given by
\begin{equation}\label{charge}
    g_{tt}^{QdS} = 1 - \frac{r_g}{r} + \frac{e^2}{r^2} -\frac{r^2}{a^2} -\left(\frac{r_q}{r}\right)^{3\omega_q + 1},
\end{equation}
where $a^2 = 3/\Lambda$. If we choose the simplest singular Schwarzschild case \cite{Liu1122,Liu1123}, the last term can be reduced to $\left(\frac{r_q}{r}\right)^{3\omega_q + 1} \longrightarrow \frac{\alpha}{r^{3\omega_q + 1}}$ just like this model. This solution (\ref{charge}) also has a rich spectrum of limits. If $e \longrightarrow 0$, there will be no charge. If $a^2 \longrightarrow \infty$, there will be no de Sitter curvature. If $r_g \longrightarrow 0, e \longrightarrow 0$, it will become a self-gravitating quintessence without black hole. Hence, if we consider the internal charge, we must start from another metric space,
\begin{equation}\label{incharge}
    g_{tt}^Q = 1 - \frac{r_g}{r} + \frac{e^2}{r^2} - \frac{\alpha}{r^{3\omega_q + 1}}.
\end{equation}

\acknowledgments
We thank the anonymous referee for quite helpful comments and suggestions.
The project is supported by the National Natural Science Foundation (No.10573004), Natural Science Foundation (NSF)
(No.10703001), Specialized Research Fund for the Doctoral Program (SRFDP)(No.20070141034) of P.R. China.

\end{document}